%% file: hcvs18_main.tex
\documentclass[copyright,creativecommons]{eptcs}
\providecommand{\event}{HCVS 2018} 

\usepackage[dvipsnames]{xcolor}
\usepackage{amsmath}
\usepackage{listings}
\usepackage{wrapfig}
\usepackage{fancyvrb}
\usepackage{enumitem}

\definecolor{gcommentcolor}{rgb}{0, 0.6, 0}

\newcommand{\nil}{[\:]}

\title{Bounded Symbolic Execution for\\Runtime Error Detection of Erlang Programs
\thanks{%
  This work has been partially supported by the EU (FEDER) and the \emph{Spanish 
  Ministerio de Ciencia, Innovaci\'on y Universidades}/AEI under grant 
  TIN2016-76843-C4-1-R, the \emph{Generalitat Valenciana} under 
  grant PROMETEO-II/2015/013 (SmartLogic), and INdAM-GNCS (Italy).
  \mbox{Emanuele~De~Angelis,} \mbox{Fabio~Fioravanti,} and Alberto~Pettorossi are research associates at CNR-IASI, Rome, Italy.  
  }
}

\author{
Emanuele De Angelis, Fabio Fioravanti
\institute{DEC, University ``G.~d'Annunzio" of Chieti-Pescara\\[1pt]
Viale Pindaro 42, 65127 Pescara, Italy}
\email{\{emanuele.deangelis, fabio.fioravanti\}@unich.it}
\and
Adri\'an Palacios\thanks{%
  Partially supported by the EU (FEDER) and the Spanish
  \emph{Ayudas para contratos predoctorales para la formaci\'on de doctores} and
  \emph{Ayudas a la movilidad predoctoral para la realizaci\'on de estancias
    breves en centros de I+D}, MINECO (SEIDI), under FPI grants BES-2014-069749
  and EEBB-I-17-12101.}
\institute{MiST, DSIC, Universitat Polit\`ecnica de Val\`encia\\[1pt]
  Camino de Vera, s/n, 46022 Val\`encia, Spain}
\email{apalacios@dsic.upv.es}
\and
Alberto Pettorossi
\institute{University of Roma Tor Vergata \\[1pt]
  Via del Politecnico 1, 00133 Roma, Italy}
\email{pettorossi@info.uniroma2.it}
\and
Maurizio Proietti
\institute{CNR-IASI\\[1pt]
  Via dei Taurini 19, 00185 Roma, Italy}
\email{maurizio.proietti@iasi.cnr.it}
}

\def\titlerunning{Bounded Symbolic Execution for Runtime Error Detection of Erlang Programs}
\def\authorrunning{E.~De Angelis et al.}

\begin{document}
\maketitle

\input{sections/0_abstract.tex}

\input{sections/1_intro.tex}

\input{sections/2_language.tex}

\input{sections/3_interpreter.tex}

\input{sections/4_conclusions.tex}

\section*{Acknowledgements}
We would like to thank the referees of HCVS 2018 for their useful and constructive 
comments. We also thank John Gallagher for stimulating discussions on the subject.


\input{sections/5_bib.bbl}
\end{document}

%% file: sections/0_abstract.tex

\begin{abstract}

Dynamically typed languages, like Erlang, allow developers to quickly write programs 
without explicitly providing any type information on expressions or function
definitions. However, this feature makes those languages less reliable than
statically typed languages, where many runtime errors can be detected 
at compile time.
In this paper, we
present a preliminary work on a tool that, by using the well-known techniques of
metaprogramming and symbolic execution, can be used to perform bounded
verification of Erlang programs.
In particular, by using Constraint Logic Programming, we develop an interpreter
that, given an Erlang program and a symbolic input for that program, returns answer
constraints that represent sets of concrete 
data for which the Erlang program generates a runtime error. 

\end{abstract}


%% file: sections/1_intro.tex
\section{Introduction}
\label{sec:intro}

Erlang~\cite{AWV96} is a functional, message passing, concurrent language with
dynamic typing.
Due to this type discipline, Erlang programmers are quite familiar with 
typing and pattern matching errors at runtime, which normally appear during the
first executions of freshly written programs. Less often, these errors will be
undetected for a long time, until the user inputs a particular value that
causes the program to crash or, in the case of concurrent programs, 
determines a particular interleaving that causes an error to occur.

In order to mitigate these problems, many static analysis tools have been
proposed.  Here let us recall:
\begin{itemize}
  \item Dialyzer~\cite{lindahl2006practical}, which is a popular tool included in
Erlang/OTP for performing type inference based on success typings, and
  \item SOTER~\cite{d2013automatic}, which is a tool that performs verification of
Erlang programs by using model checking and abstract interpretation.
\end{itemize}
However, those tools are not all fully automatic, and they can only be
used to cover either the sequential or the concurrent part of an Erlang program,
but not both.

In this paper we present a preliminary work on a technique,
based on Constraint Logic Programming (CLP)~\cite{JaM94}, for detecting
runtime errors in Erlang programs. In our approach, sequential Erlang 
programs are first translated into Prolog facts and then run by using an 
interpreter written in CLP.
Our CLP interpreter is able to run programs on symbolic input data, 
and it can perform
verification of Erlang programs up to a fixed bound on the 
number of execution steps.


%% file: sections/2_language.tex
\paragraph{The Erlang language.}
\label{sec:language}

In this work we consider only sequential programs written in a first-order
subset of the Core Erlang language\footnote{Core Erlang is
  the intermediate language used by the standard Erlang compiler, 	which
  removes most of the syntactic sugar present in Erlang programs.}. The syntax
of this subset can be found in Figure~\ref{fig:syntax}.

\begin{figure}[htb]
  \begin{center}
  $\begin{array}{rcl@{~~~~~~}l}
    \mathit{module} & ::= & \mathsf{module} ~ atom = 
    \mathit{fun}_1,\ldots,\mathit{fun}_n\\[1pt]
    \mathit{fun} & ::= & \mathit{fname} = \mathit{fundef} \\[1pt]
    {\mathit{fname}} & ::= & atom / nat\\[1pt]
    \mathit{fundef} & ::= & \mathsf{fun}~(X_1,\ldots,X_n)~\texttt{->}~expr~\mathsf{end}\\
    lit & ::= & atom ~\mid~ int ~\mid~ \nil \\[1pt]
    expr & ::= & \mathit{X} ~\mid~ lit ~\mid~ \mathit{fname} 
    ~\mid~ [\,expr_1\,|\,expr_2\,] ~\mid~ \{\,expr_1,\ldots,expr_n\,\} \\[1pt]
    & \mid &
    \mathsf{let}~~\mathit{X}=expr_1~~\mathsf{in}~~expr_2 \\[1pt]
    & \mid &
    \mathsf{case}~~expr~~\mathsf{of}~~clause_1;\ldots;clause_m~~\mathsf{end}\\[1pt]
    & \mid &
    \mathsf{apply}~~fname~(~expr_1,\ldots,expr_n~) \\ [1pt]
    & \mid &
    \mathsf{call}~~atom:fname~(~expr_1,\ldots,expr_n~) \\[1pt]
    & \mid &
    \mathsf{primop}~~atom~(~expr_1,\ldots,expr_n~) \\ [1pt]
    & \mid &
    \mathsf{try}~~expr_1~~\mathsf{of}~~X_1~~\texttt{->}~~expr_2~~\mathsf{catch}~~X_2~~\texttt{->}~~expr_3 \\
    clause & ::= & pat ~~\mathsf{when}~~expr_1~~\texttt{->}~~expr_2 \\[1pt]
    pat & ::= & \mathit{X} ~\mid~ lit ~\mid~ [\,pat_1\,|\,pat_2\,] ~\mid~
    \{\,pat_1,\ldots,pat_n\,\} \\[1pt]
  \end{array}
  $
  \end{center}
\caption{Language syntax rules}
\label{fig:syntax}
\vspace{2pt}
\end{figure}

Here, a module is a sequence of function declarations, where each function name has an associated definition of the form
``$\mathsf{fun}~(X_1,\ldots,X_n)~\texttt{->}~expr~\mathsf{end}$'' 
(for simplicity, we assume that programs are made out of a single module).  
The body of a function is an expression \emph{expr}, which can include 
variables, 
literals (atoms, integers, floats, or the empty list), 
list constructors, 
tuples, 
let expressions, 
case expressions, 
function applications,  
calls to built-in functions, and
try/catch blocks.

In a case expression
``$\mathsf{case}~expr~\mathsf{of}~clause_1;\ldots;clause_m~\mathsf{end}$'', the
expression $expr$ is first reduced to a value $v$, and then $v$ is matched
against the clauses ``$pat ~\mathsf{when}~expr_1~\texttt{->}~expr_2$'' of the
case expression. The first clause to match this value (i.e., the first clause for which
there exists a
substitution $\sigma$ such that {$v = pat\,\sigma$ and $expr_1\,\sigma$} reduces to
$\mathsf{true}$) is selected, and the computation continues with the evaluation
of the clause body (after updating the environment with $\sigma$).

Let us remark that primop expressions of the form
``$\mathsf{primop}~atom~(expr_1,\ldots,expr_n)$'' are primitive operation calls.
In general, their evaluation is implementation-dependent, and they may have side
effects or raise exceptions. However, in our setting, these are mainly used for
raising exceptions in pattern matching errors.

The Erlang program in Figure~\ref{fig:run_example} will successfully compile with 
no warnings in Erlang/OTP and will correctly compute the sum of the elements in
\texttt{L} provided that \texttt{L} is a list of numbers. 

\setlength{\columnsep}{20pt}%
\begin{wrapfigure}{L}{6cm}
\vspace{-5pt}
\begin{verbatim}
  -module(sum_list).
  -export([sum/1]).
    
  sum(L) ->
    case L of
      [] -> 0;
      [H|T] -> H + sum(T)
    end.
\end{verbatim}
\vspace{-5pt}
\caption{A program in Erlang that computes the sum 
of all numbers in the input list \texttt{L}.}
\label{fig:run_example}
\vspace{-10pt}
\end{wrapfigure}

Otherwise, the program 
generates a runtime error. For instance,
if the input to \texttt{sum} is an atom, then the program crashes and 
outputs a pattern matching error (\texttt{match\_fail}),
because there are no patterns that match an atom.
Similarly, if the input to \texttt{sum} is a list of values, where at least one element 
is an atom, the execution halts with
a type error (\texttt{badarith}), when applying the function `\texttt{+}' to a 
non-numerical argument.

The tool Dialyzer does not generate any warnings when analyzing this program. That
is coherent with the Dialyze approach, which only complains about type
errors that would guarantee the program to crash for all input values. However, it might
be the case that we want to perform a more detailed analysis on this
program. In the following, we will see how our tool lists all the
potential runtime errors, together with the input types that can cause
them. 


%% file: sections/3_interpreter.tex
\section{Symbolic Interpreter for Runtime Error Detection of Erlang Programs}
\label{sec:interpreter}

The main component of the verifier is a CLP interpreter that defines the operational
semantics of our language. 
This executable specification of the semantics enables the execution of Erlang programs
represented as Prolog facts.

Therefore, we have defined a translation from Erlang programs to Prolog facts.
More precisely, our translator generates one fact \texttt{fundef} for each function
definition occurring in the Core Erlang code obtained from the original Erlang module.
For instance, the translation of the function \texttt{sum/1} defined in the module
\texttt{sum\_list} corresponds to the fact  which can be seen in 
Figure~\ref{fig:translation}.

\begin{figure}[htb]
\noindent
\texttt{
fundef(lit(atom,'sum\_list'),var('sum',1),\\
\hspace*{3ex}fun([var('@c0')],\\
\hspace*{5ex}case(var('@c0'),[\\
\hspace*{7ex}clause([lit(list,nil)],lit(atom,'true'),\\
\hspace*{9ex}lit(int,0)),\\
\hspace*{7ex}clause([cons(var('H'),var('T'))],lit(atom,'true'),\\
\hspace*{9ex}let([var('@c1')],apply(var('sum',1),[var('T')]),\\
\hspace*{11ex}call(lit(atom,'erlang'),lit(atom,'+'),[var('H'),var('@c1')]))),\\
\hspace*{7ex}clause([var('@c2')],lit(atom,'true'),\\
\hspace*{9ex}primop(lit(atom,'match\_fail'),\\
\hspace*{11ex}[tuple([lit(atom,'case\_clause'),var('@c2')])]))]))).
}
\caption{Prolog fact generated by the Erlang-to-Prolog translation for the \texttt{sum} function
  definition.}
\label{fig:translation}
\end{figure}

This translation is quite straightforward, since the standard compilation from
Erlang to Core Erlang greatly simplifies the code. Note that, since we generate
Prolog facts, we have used the \texttt{cons} predicate for Erlang list
constructors to distinguish them from Prolog list constructors. Note also that
an additional \emph{catch-all} clause has been added for the case in which the
argument does not match any of the clauses from the case expression (i.e., for
pattern matching errors). This transformation and similar ones are automatically
made by the standard compilation from Erlang to Core Erlang.

The CLP interpreter provides a flexible means to perform the \emph{bounded
verification} of Erlang programs. By using   
a symbolic representation
of the input data, the interpreter allows the exhaustive exploration of the
program computations without explicitly enumerating all the concrete input
values.  In particular, the interpreter can run on input terms containing
logic variables, and it uses constraint solvers to manipulate expressions with
variables ranging over integer or real numbers. By fixing a bound to limit the
number of computation steps performed by the interpreter, we force the
exploration process to be finite, and hence either we detect a runtime error or
we prove that the program is error-free up to the given bound. 

Let us consider an Erlang program \texttt{Prog}, which is translated into Prolog 
facts as shown in Figure~\ref{fig:translation}. 
In order to perform the bounded verification of the Erlang program \texttt{Prog},
the interpreter provides the predicate \texttt{run(FName/Arity,Bound,In,Out)},
whose execution evaluates the application of the function \texttt{FName} 
of arity~\texttt{Arity} to the input arguments \texttt{In} in at most \texttt{Bound}
steps.
The arguments \texttt{In} are represented as a Prolog list (written using square 
brackets) of length \texttt{Arity}.
\texttt{Out} is the result of the function application.
If the evaluation of the function application generates an error, 
then \texttt{Out} is bound to a term of the form \texttt{error(Err)}, 
where \texttt{Err} is an error name (e.g., \texttt{match\_fail}, indicating a match
failure, or \texttt{badarith}, indicating an attempt to evaluate an arithmetic function
on a non-arithmetic input), meaning that the specific error \texttt{Err} had occurred.
Hence, the bounded verification of a given Erlang program can be performed by executing 
a query of the form:

\medskip

\texttt{?- run(FName/Arity,Bound,In,error(Err)).}

\medskip
\noindent
where \texttt{FName} is a constant,  \texttt{Arity} and \texttt{Bound} are
non-negative integers, and \texttt{In} and \texttt{Err}  are, possibly non-ground,
terms.

Any answer to the query is a successful detection of the error \texttt{Err} 
generated by evaluating the application of the function \texttt{FName} 
to the input~\texttt{In}. 
If no answer is found, then it means that no error is generated by exploring the
computation of \texttt{FName} up to the value of \texttt{Bound}, and we say that 
the program \texttt{Prog}  is correct up to the given bound.

Now let us see the bounded verifier in action by considering the 
\texttt{sum\_list} program of the previous section and the following query:

\medskip

\texttt{?- run(sum/1,20,In,error(Err)).}

\medskip
\noindent
Among the answers to the query, we get the following constraints relative to
 the input \texttt{In} and the error~\texttt{Err}:

\medskip
\texttt{In=[cons(lit(Type,\_V),lit(list,nil))],}
\noindent

\texttt{Err=badarith,}
\noindent

\texttt{dif(Type,int), dif(Type,float)}

\medskip
\noindent
meaning that if \texttt{sum/1} takes as input a list (represented as a Prolog 
term of the form \texttt{cons(Head,Tail)}) whose head is not a numeric literal 
(denoted by the constraints \texttt{dif(Type,int)} and \texttt{dif(Type,float)}), 
then a \texttt{badarith} error occurs, that is, a non-numerical argument is given 
as input to an arithmetic operator. Another answer we get is:

\medskip
\texttt{In=[L],}\nopagebreak

\texttt{Err=match\_fail,}\nopagebreak

\texttt{dif(L,cons(\_Head,\_Tail)), dif(L,lit(list,nil))}

\medskip
\noindent
meaning that if \texttt{sum/1} takes as input an argument  \texttt{L}
which is neither a \texttt{cons} nor a \texttt{nil} term, then a
\texttt{match\_fail} error occurs.
Note that, due to the recursive definition of \texttt{sum}, the bound is essential to
detect this error.

Now let us  introduce the predicate \texttt{int\_list(L,M)} that generates lists
\texttt{L} of integers of length up to~\texttt{M}. 
For instance, the query 

\medskip
\texttt{?- int\_list(L,100).}

\medskip
\noindent	
generates the answer 

\smallskip
\texttt{L=cons(lit(int,N1),cons(lit(int,N2),...))}

\medskip
\noindent
where \texttt{L} is a list of length 100 and $\texttt{N1}, \texttt{N2},\ldots, \texttt{N100}$ are variables.
If we give \texttt{L} as input to \texttt{sum} as follows:

\medskip
\texttt{?- int\_list(L,100), run(sum/1,100,L,error(Err)).}

\medskip
\noindent
the bounded verifier terminates after 0.347 seconds\footnote{The query has been executed
using SWI-Prolog v7.6.4 (\url{http://www.swi-prolog.org/}) on an Intel Core i5-2467M 1.60GHz processor with 4GB of memory under GNU/Linux OS} 
with answer \texttt{false}, meaning that if the input to \texttt{sum} is any list of 100
integers, then the program is correct up to the bound \texttt{100}.
Note that no concrete integer element of the list is needed for the verification of this
property.

Now we sketch the implementation of the operational semantics of Erlang
expressions.  The semantics is given in terms of a transition relation of the
form \texttt{tr(Bound,ICfg,FCfg)}, which defines how to get the final
configuration \texttt{FCfg} from the initial configuration \texttt{ICfg} in
\texttt{Bound} computation steps.
Configurations are pairs of the form
\texttt{cf(Env,Exp)}, where \texttt{Env} is the environment mapping program
variables to values and \texttt{Exp} is a term representing an Erlang
expression.

The environment is extended with a boolean flag that keeps track of the occurrence of 
any runtime error during program execution.
The value of the error flag \texttt{Flag} in the environment \texttt{Env} can be 
retrieved by using the predicate \texttt{lookup\_error\_} \texttt{flag(Env,Flag)}. 
The value of the flag in a given environment~\texttt{EnvIn} can be updated by using the
predicate \texttt{update\_error\_flag(EnvIn,} 
\texttt{Flag,EnvOut)}, 
thereby deriving the environment \texttt{EnvOut} whose error flag is set to \texttt{Flag}.
If the evaluation of \texttt{IExp} generates the error \texttt{Err}, 
then \texttt{FExp} is a term of the form \texttt{error(Err)} and the error flag is 
set to \texttt{true}.

In Figure~\ref{fig:applytr} we present the clause for \texttt{tr/3} that
implements the semantics of function applications represented using terms of the
form \texttt{apply(FName/Arity,IExps)}, where \texttt{FName} is the name of a
function of arity \texttt{Arity} applied to the expressions \texttt{IExps}.
The transition {performed by \texttt{tr/3}} only applies if:

\begin{enumerate}
	\item no error has occurred so far, that is, \texttt{lookup\_error\_flag(IEnv,false)}, and
	\item the bound has not been exceeded, that is, \texttt{Bound\,>\,0}. 
\end{enumerate}


\begin{figure} 

\hspace*{9mm}\begin{Verbatim}[baselinestretch=1.1]
            tr(Bound,cf(IEnv,IExp),cf(FEnv,FExp)) :-
                  IExp = apply(FName/Arity,IExps),
                  lookup_error_flag(IEnv,false),            % 1
                  Bound > 0,                                % 2
                  Bound1 is Bound - 1,                      % 3
                  lookup_fun(FName/Arity,FPars,FBody),      % 4
                  tr_list(Bound1,IEnv,IExps,EEnv,EExps),    % 5
                  bind(FPars,EExps,AEnv),                   % 6
                  lookup_error_flag(EEnv,F1),
                  update_error_flag(AEnv,F1,BEnv),          % 7
                  tr(Bound1,cf(BEnv,FBody),cf(CEnv,FExp)),  % 8
                  lookup_error_flag(CEnv,F2), 
                  update_error_flag(EEnv,F2,FEnv).          % 9
\end{Verbatim}
\vspace{-5pt}
\caption{Definition of the operational semantics for a function application \texttt{apply/2}.}
\label{fig:applytr}
\end{figure}  

\smallskip
\noindent
Then, 
the following operations are performed:

%
%
%

\begin{enumerate}[resume]
  \item the value of the bound \texttt{Bound} is decreased,
  
  \item the parameters \texttt{FPars} and the body \texttt{FBody} of the function
  \texttt{FName} of arity \texttt{Arity} are retrieved 
    (the predicate \texttt{lookup\_fun/3} is responsible for extracting \texttt{FPars}
    and \texttt{FBody} from the \texttt{fundef} fact representing \texttt{FName/Arity}),
   
  \item the list of the actual parameters \texttt{IExps} is evaluated in
    \texttt{IEnv}, thereby deriving the list of expressions \texttt{EExps} and
    the new environment \texttt{EEnv} (it may differ from \texttt{IEnv} in the
    error flag and new variables occurring in the expressions \texttt{IExps} may
    have been added),
    
  \item the formal parameters \texttt{FPars} are bound to the expressions
     \texttt{IExps}, thereby deriving the new environment \texttt{AEnv}, 
     
  \item the error flag in \texttt{AEnv} is updated to the value \texttt{F1} {in} \texttt{EEnv}, thereby deriving the environment \texttt{BEnv},
    
  \item the body \texttt{FBody} is evaluated in \texttt{BEnv} to get the final
    expression \texttt{FExp}, and
    
  \item the final environment \texttt{FEnv} is obtained from \texttt{EEnv} 
    by setting the error flag to the value \texttt{F2} obtained from the callee function.
    
\end{enumerate}


Each rule of the operational semantics for Erlang programs is translated into a clause
for the predicate  \texttt{tr/3}. These clauses are omitted. 

Now we can present the definition of \texttt{run/4}, which depends on \texttt{tr/3}:

\begin{verbatim}
 run(FName/Arity,Bound,In,Out) :-
   lookup_fun_pars(FName/Arity,FPars),
   bind(FPars,In,IEnv),
   tr(Bound,cf(IEnv,apply(FName/Arity,FPars)),cf(FEnv,Out)).
\end{verbatim}

\medskip
The predicate \texttt{run} retrieves the formal parameters \texttt{FPars} of
\texttt{FName/Arity} and creates an environment \texttt{IEnv} where those parameters are
bound to the input values \texttt{In}. 
Then, it evaluates the application of \texttt{FName} to its parameters, thereby
producing the final expression \texttt{Out}.


%% file: sections/4_conclusions.tex
\section{Conclusions and future work}
\label{sec:conclusions}

We have presented a work in progress for the development of a CLP interpreter for
detecting runtime errors of Erlang programs. An Erlang program is first translated
into a set of Prolog facts, then the CLP interpreter is run using this translation
together with symbolic input data.  At present, our interpreter
is able to deal with first-order sequential Erlang programs, but we
think that the extension to higher-order functions can be achieved by
following a similar approach. In the future, we also plan to consider
concurrency with an appropriate technique for handling the state
explosion problem. 
For instance, we may employ a partial order reduction technique~\cite{abdulla2014optimal}
to obtain the minimal set of concurrent behaviours for a given program, and 
then generate the associated executions using our interpreter.

Let us briefly compare our work with the static analysis tools available for Erlang.
Unlike Dyalizer~\cite{lindahl2006practical}, our tool computes answer constraints that
describe type-related input patterns which lead to runtime errors. 
However, as already mentioned, due to the bounded symbolic execution, our interpreter may
terminate with no answer, even if runtime errors are possible for concrete runs which
exceed the given bound. 
One of the weaknesses of Dialyzer is that it is hard to know where typing errors 
come from.
An extension of Dialyzer that provides an explanation for the cause of typing errors has
been proposed to overcome this problem~\cite{sagonas2013precise}. 
We believe that we are able to provide a similar information if we include debugging
information in the clauses generated by our Erlang-to-CLP translation.

Unlike SOTER~\cite{d2013automatic}, which is based on abstract interpretation, our CLP
interpreter provides full support to arithmetic operations through the use of constraint
solvers. 
Moreover, the symbolic interpreter does not require any user intervention (except for the
bound), while in SOTER the user is responsible for providing a suitable abstraction.

Besides being useful on its own for bounded verification, the CLP interpreter for Erlang
may be the basis for more sophisticated analysis techniques.
In particular, by following an approach developed in the case of imperative languages,
we intend to apply CLP transformation techniques to specialize the interpreter with
respect to a given Erlang program and its symbolic input~\cite{de2017semantics}.
The specialized CLP clauses may enable more efficient bounded verification, and they 
can also be used as input to other tools for analysis and verification (such as
constraint-based analyzers~\cite{De&14b,Ka&16} and SMT solvers~\cite{Ho&12,DeB08}), which
have already been shown to be effective in other contexts~\cite{De&14c,De&15d,De&17c}.
Moreover, the specialized clauses can be used to apply backward analysis techniques for
CLP programs based on abstract interpretation (see, for instance, \cite{Ho&04,Ka&18}). 
{Backward analysis aims at deriving from a property that is expected to hold at the
end of the execution of a program, conditions on the query which guarantee that the desired property indeed holds.}
In our context, backward analysis can be applied to deduce those conditions that may 
cause a runtime error, and then use them to improve the forward symbolic execution of the
program.
